\documentclass[
]{ceurart}

\usepackage{listings}
\usepackage{todonotes}
\usepackage{enumitem}
\definecolor{link}{RGB}{0, 123, 255}

\begin{document}

\copyrightyear{2026}
\copyrightclause{Copyright for this paper by its authors. Use permitted under Creative Commons License Attribution 4.0 International (CC BY 4.0).}

\conference{Joint Proceedings of REFSQ-2026 Workshops, Doctoral Symposium, Posters \& Tools Track, and Education and Training Track. Co-located with REFSQ 2026. Poznan, Poland, March 23-26, 2026}

\title{EmpiRE-Compass: A Neuro-Symbolic Dashboard for Sustainable and Dynamic Knowledge Exploration, Synthesis, and Reuse}
\subtitle{Navigating the Knowledge Landscape of Empirical Research in Requirements Engineering}

%\metatitle{EmpiRE-Compass: A Neuro-Symbolic Dashboard for Sustainable and Dynamic Knowledge Exploration, Synthesis, and Reuse}

%\tnotetext[1]{You can use this document as the template for preparing your publication. We recommend using the latest version of the ceurart style.}

% \metaauthor*{\uri{https://orcid.org/0000-0001-5336-6899}{Oliver Karras}}
% \metaauthor*{\uri{https://orcid.org/0009-0002-1165-773X}{Amirreza Alasti}}
% \metaauthor*{\uri{https://orcid.org/0009-0007-2097-9761}{Lena John}}
% \metaauthor*{Sushant Aggarwal}
% \metaauthor*{Yücel Celik}

\author[1]{Oliver Karras}[%
orcid=0000-0001-5336-6899,
email=oliver.karras@tib.eu,
]
\cormark[1]

\author[2]{Amirreza Alasti}[%
orcid=0009-0002-1165-773X,
email=amirreza.alasti@stud.uni-hannover.de,
]

\author[1]{Lena John}[%
orcid=0009-0007-2097-9761,
email=lena.john@tib.eu,
]

\author[2]{Sushant Aggarwal}[%
orcid=,
email=sushant.aggarwal@stud.uni-hannover.de,
]

\author[2]{Yücel Celik}[%
orcid=,
email=yucel.celik@stud.uni-hannover.de,
]

\address[1]{TIB - Leibniz Information Centre for Science and Technology, Germany}
\address[2]{Leibniz University Hannover, Germany}

\cortext[1]{Corresponding author.}
%\fntext[1]{These authors contributed equally.}

\begin{abstract}
\textit{\textbf{[Background.]}} Software engineering (SE) and requirements engineering (RE) face a significant increase in secondary studies, particularly literature reviews (LRs), due to the ever-growing number of scientific publications. Generative artificial intelligence (GenAI) exacerbates this trend by producing LRs rapidly but often at the expense of quality, rigor, and transparency. At the same time, secondary studies often fail to share underlying data and artifacts, limiting replication and reuse.
\textbf{[Objective.]} This paper introduces \href{https://empire-compass.tib.eu/R186491/}{\textit{\textcolor{link}{EmpiRE-Compass}}}, a neuro‑symbolic dashboard designed to lower barriers for accessing, replicating, and reusing LR data. Its overarching goal is to demonstrate how LRs can become more sustainable by semantically structuring their underlying data in research knowledge graphs (RKGs) and by leveraging large language models (LLMs) for easy and dynamic access, replication, and reuse.
\textbf{[Method.]} Building on two RE use cases, we developed \href{https://empire-compass.tib.eu/R186491/}{\textit{\textcolor{link}{EmpiRE-Compass}}} with a modular system design and workflows for curated and custom competency questions.
The dashboard is freely available online, accompanied by a \href{https://doi.org/10.5446/72249}{\textit{\textcolor{link}{demonstration video}}}. To manage operational costs, a limit of 25 requests per IP address per day applies to the default LLM (GPT-4o mini). All source code and documentation are released as an \href{https://github.com/okarras/EmpiRE-Compass}{\textit{\textcolor{link}{open‑source project}}} to foster reuse, adoption, and extension.
\textbf{[Results.]} \mbox{\href{https://empire-compass.tib.eu/R186491/}{\textit{\textcolor{link}{EmpiRE-Compass}}}} provides three core capabilities: (1) Exploratory visual analytics with visualizations, interpretations, and explanations for curated competency questions; (2) Neuro‑symbolic synthesis for custom competency questions with dynamic visualizations, interpretations, explanations, and direct access to underlying RKG data; and (3) Reusable knowledge with all queries, analyses, and results openly available.
\textbf{[Conclusions.]} By unifying RKGs and LLMs in a neuro‑symbolic dashboard, \href{https://empire-compass.tib.eu/R186491/}{\textit{\textcolor{link}{EmpiRE-Compass}}} advances sustainable LRs in RE, SE, and beyond. It lowers technical barriers, fosters transparency and reproducibility, and enables collaborative, continuously updated, and reusable~LRs.
\end{abstract}

% \method*{Development of EmpiRE-Compass: A neuro‑symbolic dashboard designed to lower barriers for accessing, replicating, and reusing LR data.}
% \conclusion*{EmpiRE-Compass advances sustainable LRs in RE, SE, and beyond}
% \contribution*{paper class}{Tools paper}
% \contribution*{replication package}{http://dx.doi.org/10.5281/zenodo.18170203}
% \contribution*{code repository}{https://github.com/okarras/EmpiRE-Compass}
% \contribution*{dataset}{https://orkg.org/observatories/Empirical_Software_Engineering}
% \contribution*{dataset}{https://orkg.org/observatories/nlp4re}
% \researchfield*{\uri{https://orkg.org/resource/R215741}{Requirements Engineering}}

\maketitle

%https://2026.refsq.org/track/refsq-2026-posters---tools#Call-for-Submissions

\section{Introduction}
Software engineering (SE), including requirements engineering (RE), faces a significant increase in secondary studies, particularly literature reviews (LRs), due to the ever-growing number of scientific publications~\cite{Napoleao.2022, Auer.2023, Karras.2025}. The rise of generative artificial intelligence (GenAI) has exacerbated this problem, as it is increasingly used to generate LRs rapidly and automatically, but often at the expense of quality, rigor, and transparency -- a trend so concerning that major preprint servers have changed their policies\footnote{arXiv banned non-peer-reviewed LRs from computer science to counteract the flood of low-quality, AI-generated content~\cite{Castelvecchi.2025}.}~\cite{Castelvecchi.2025}. 

The challenge of LRs, however, goes beyond the recent GenAI trend: The core problem is that extracted and analyzed data, along with the corresponding research artifacts, are often unavailable~\cite{Karras.2023}. This lack of underlying data and artifacts poses a substantial obstacle to collaboration and updating of LRs, as the analyses and results cannot be replicated and reused~\cite{Karras.2025}. A recent systematic mapping study found that secondary studies in SE, published between 2013 and 2023, are increasingly adopting open science practices. However, only 31.5\% of the 537 analyzed studies shared their data and artifacts, and just 12.1\% deposited them in a persistent repository~\cite{Huotala.2025}. Although openly available data and artifacts in persistent repositories represent progress toward the broader vision of open science in SE~\cite{Huotala.2025}, current practices still fail to leverage the full technological potential to support replication and reuse effectively~\cite{Karras.2025, Karras.2023, DosSantos.2024, John.2025}. It is necessary to move beyond static, file-based representations in repositories by using semantically structured representations integrated into open science infrastructures aligned with the FAIR data principles~\cite{Napoleao.2022, DosSantos.2024, Wilkinson.2016}. In this way, data and artifacts become interpretable and actionable for humans and machines, enabling \textit{sustainable} exploration, synthesis, and reuse~\cite{Karras.2025, John.2025}.

Building on this vision, we used the \href{https://orkg.org/}{\textit{\textcolor{link}{Open Research Knowledge Graph (ORKG)}}}~\cite{Auer.2023, Auer.2025} to demonstrate how data from an LR with 776 papers on the state and evolution of empirical research in RE can be semantically structured in an open science infrastructure to improve its availability for replication and reuse~\cite{Karras.2023}. The resulting research knowledge graph (RKG), called \href{https://orkg.org/observatories/Empirical_Software_Engineering}{\textit{\textcolor{link}{KG-EmpiRE}}}, and its \href{https://github.com/okarras/EmpiRE-Analysis}{\textit{\textcolor{link}{analysis script}}} constitute long-term available and reusable research artifacts\footnote{Recognized at \href{https://conf.researchr.org/track/RE-2024/RE-2024-artifacts}{\textit{\textcolor{link}{RE’24 Artifact Track}}} with the badges ``Available'' and ``Reusable'' as well as the \href{https://www.oliver-karras.de/wp-content/uploads/2024/06/IEEE_RE2024_Certifiacte_Best_Artifact_Award.pdf}{\textit{\textcolor{link}{Best Artifact Award}}}.}~\cite{Karras.2024b, Karras.2024d}. Based on these results, we advocated for the broader use of RKGs for sustainable LRs in SE and beyond~\cite{Karras.2025}. However, while RKGs ensure that LR data is openly available and semantically structured, their effective use demands advanced technical skills, such as writing SPARQL queries or navigating complex graph schemas, to represent, query, analyze, and interpret the data. Many researchers lack these skills and thus face barriers that continue to impede access, replication, and reuse.

To address these barriers, we build on the established use of dashboards to enable intuitive, human‑centered access. While dashboards support exploration, synthesis, and reuse, they typically provide predefined visualizations with only minor customization that users must interpret themselves~\cite{Alhamadi.2022}. We go beyond these limitations by
integrating a neuro‑symbolic approach that leverages the synergy between large language models (LLMs) and RKGs. This approach unifies the neural, generative capabilities of LLMs with symbolic representations of semantically structured data in RKGs~\cite{Pan.2024}. In particular, LLMs can \textit{dynamically} generate custom visualizations, interpretations, and explanations directly from RKG data as answers to competency questions\footnote{A competency question is a natural language query expressing an information need that an RKG must be able to answer.}. This synergy enables researchers to interact with RKGs easily and flexibly, lowering the barriers to access, replication, and reuse.

We applied this idea to our KG-EmpiRE use case~\cite{Karras.2023}, resulting in \href{https://empire-compass.tib.eu/R186491/}{\textit{\textcolor{link}{EmpiRE-Compass}}}: A neuro‑symbolic dashboard that allows researchers to explore, synthesize, and reuse knowledge about empirical research in RE \textit{sustainably} and \textit{dynamically}. \href{https://empire-compass.tib.eu/R186491/}{\textit{\textcolor{link}{EmpiRE-Compass}}} offers three core capabilities: (1) \textit{Exploratory visual analytics}, investigating curated competency questions using predefined and dynamic visualizations, interpretations, and explanations; (2) \textit{Neuro‑symbolic synthesis}, combining RKGs and LLMs to answer custom competency questions with dynamically generated visualizations, interpretations, explanations, and direct access to the underlying data; and (3) \textit{Reusable knowledge}, making all structured data, queries, analyses, and results openly available for replication, reuse, and sharing. Beyond KG‑EmpiRE, we already extended the dashboard with a second use case on \href{https://empire-compass.tib.eu/R1544125}{\textit{\textcolor{link}{empirical research in NLP4RE}}}~\cite{Alasti.2026} based on 50 \href{https://zenodo.org/records/14197338}{\textit{\textcolor{link}{NLP4RE ID Cards}}}~\cite{Abualhaija.2024}, which are organized in the \href{https://doi.org/10.48366/R1584378}{\textit{\textcolor{link}{ORKG}}}~\cite{Karras.2025a}. We make the following contributions:

\begin{enumerate}[noitemsep]
\vspace{-0.2cm}
    \item Presenting \href{https://empire-compass.tib.eu/R186491/}{\textit{\textcolor{link}{EmpiRE-Compass}}}, along with its system design and workflows, incorporating both use cases (KG‑EmpiRE and NLP4RE ID Card) and demonstrating the integration of a neuro‑symbolic approach into a dashboard for LR data.
    \item Releasing the complete source code and documentation as \href{https://github.com/okarras/EmpiRE-Compass}{\textit{\textcolor{link}{open‑source project}}}~\cite{Karras.2026} to foster reuse, adoption, and extension within the research community.
\vspace{-0.2cm}
\end{enumerate}

\section{Related Work}
Dashboards are established tools for making complex data available and accessible across domains~\cite{John.2025a}. We identified 8 dashboards\footnote{Two dashboards are no longer accessible. Links to the remaining six dashboards are included in the listed disciplines.} developed to support exploration, synthesis, and reuse of LR data in various disciplines, such as environmental science~\cite{Scheuer.2021}, \href{https://softwarearchitectureresearch.github.io/visulite/}{\textit{\textcolor{link}{software architecture}}}~\cite{Kaplan.2025}, \href{http://openbiodiv.net/literature-exploration?type=taxons}{\textit{\textcolor{link}{biodiversity}}}~\cite{Penev.2019}, \href{https://covid-aqs.fz-juelich.de/}{\textit{\textcolor{link}{atmospheric science}}}~\cite{Gkatzelis.2021}, \href{https://app.cooperationdatabank.org/}{\textit{\textcolor{link}{social science}}}~\cite{Spadaro.2022}, educational science~\cite{Lezhnina.2022}, \href{https://hypothesis-evidence-explorer.hi-knowledge.org/}{\textit{\textcolor{link}{invasion biology}}}~\cite{Bernard-Verdier.2024}, and \href{https://orkg.org/usecases/r0-estimatesovid-dashboard}{\textit{\textcolor{link}{virology}}}~\cite{Shamsabadi.2024}.

Among these eight dashboards, two approaches~\cite{Scheuer.2021, Kaplan.2025} rely on spreadsheets, three~\cite{Penev.2019, Gkatzelis.2021, Spadaro.2022} use different RKGs, and three~\cite{Lezhnina.2022, Shamsabadi.2024, Bernard-Verdier.2024} build directly on the ORKG. Kaplan et al.~\cite{Kaplan.2025}, while currently using spreadsheets, plan to migrate their data into the ORKG. The prominent use of RKGs reflects the envisioned broader transition toward semantically structured representations of LR data integrated into open science infrastructures aligned with the FAIR data principles.

Although these dashboards demonstrate clear benefits for improving data availability, accessibility, and transparency, they remain limited to predefined visualizations with only minor customization and provide no interpretations or explanations to users. Our work builds on these efforts and benefits, but goes further by enhancing exploration, synthesis, and reuse through unification of LLMs and RKGs -- combining semantically structured data with flexible and intuitive user interaction using natural language. To the best of our knowledge, dashboards for LR data have not previously employed a neuro‑symbolic approach. \href{https://empire-compass.tib.eu/R186491/}{\textit{\textcolor{link}{EmpiRE-Compass}}} demonstrates how this integration enables sustainable and dynamic knowledge exploration, synthesis, and reuse.

%Important aspects:
%- Describe the goals, requirements, intended users, and context of the use of the tool.
%- Clearly explain the architecture and inner workings of the tool (at least two pages).
%- Detail the workflow or interaction of a user with the tool (half a page at least).
\section{The Neuro-Symbolic Dashboard: EmpiRE-Compass}
\begin{figure}[htbp]
    \vspace{-0.1cm}
    \centering
    \includegraphics[width=0.9\textwidth]{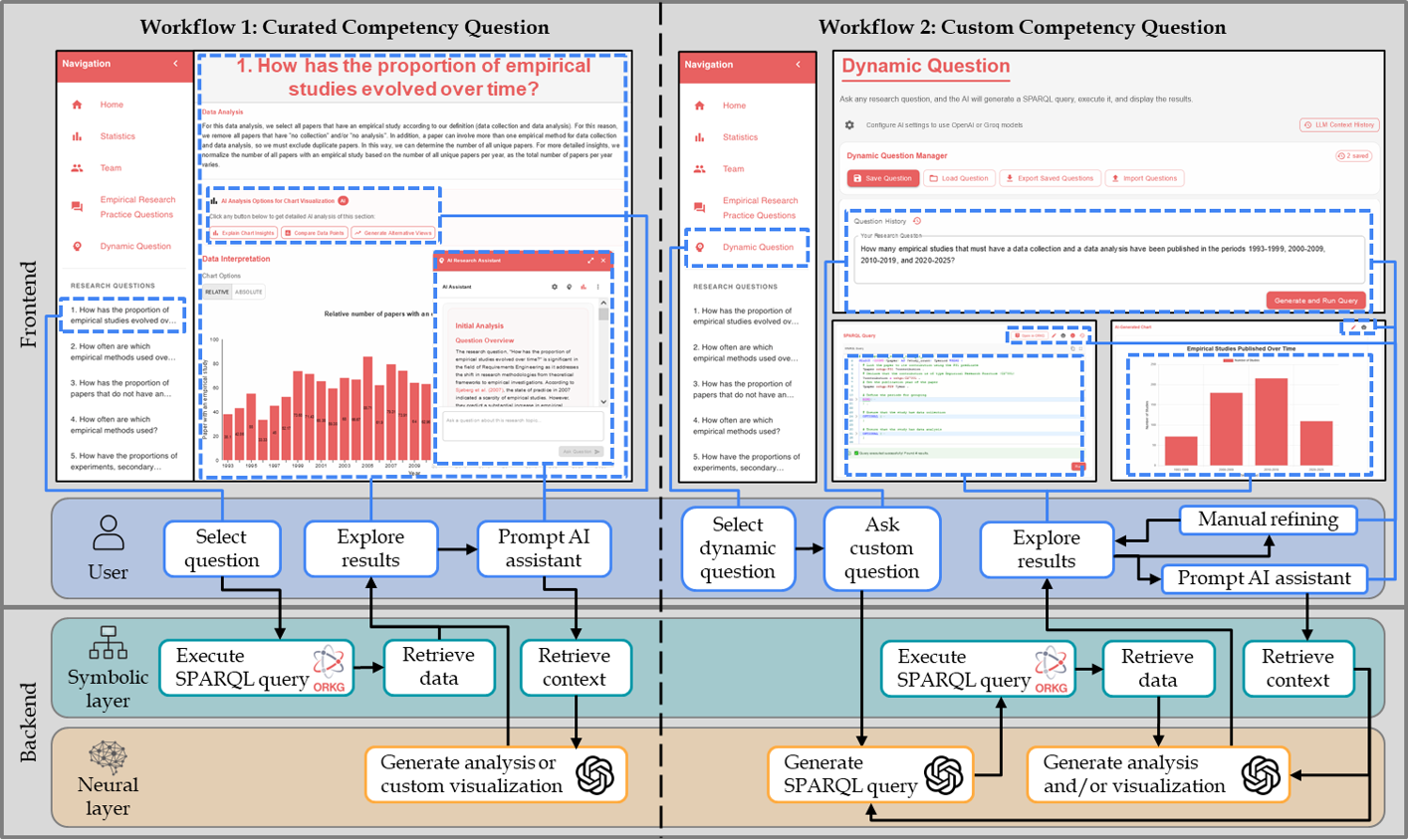}
    \caption{System design of \href{https://empire-compass.tib.eu/R186491/}{\textit{\textcolor{link}{EmpiRE-Compass}}} with its two main workflows for knowledge exploration, synthesis, and reuse based on curated (\href{https://empire-compass.tib.eu/R186491/questions/1}{\textit{\textcolor{link}{Workflow 1}}}) and custom (\href{https://empire-compass.tib.eu/R186491/dynamic-question}{\textit{\textcolor{link}{Workflow 2}}}) competency questions\protect\footnotemark.}
    \label{fig:architecture_workflows}
    \vspace{-0.9cm}
\end{figure}

\href{https://empire-compass.tib.eu/R186491/}{\textit{\textcolor{link}{EmpiRE-Compass}}} is a progressive web application designed to support researchers in exploring, synthesizing, and reusing knowledge about empirical research in RE through curated and custom competency questions. Its overarching goal is to demonstrate how LRs can become more sustainable by semantically structuring their underlying data in RKGs and by leveraging LLMs for easy and dynamic access. This neuro-symbolic approach enhances availability, accessibility, and transparency for replication and reuse. \figurename{~\ref{fig:architecture_workflows}} illustrates the system design and the two main workflows, explained below.

\footnotetext{The custom competency question used is provided online as example question ``\href{https://empire-compass.tib.eu/R186491/dynamic-question}{\textit{\textcolor{link}{Number of empirical studies per decade}}}''.}

\subsection{System Design: Requirements, Architecture, and Implementation}
\href{https://empire-compass.tib.eu/R186491/}{\textit{\textcolor{link}{EmpiRE-Compass}}} is designed to meet a set of functional and non‑functional requirements derived from its intended use. Functionally, the dashboard must \hypertarget{FR1}{(FR1) support curated competency questions}, \hypertarget{FR2}{(FR2) allow custom natural language competency questions}, \hypertarget{FR3}{(FR3) execute SPARQL queries against the ORKG}, \hypertarget{FR4}{(FR4) process and visualize retrieved data}, \hypertarget{FR5}{(FR5) provide contextual interpretations and explanations}, \hypertarget{FR6}{(FR6) enable manual and prompt‑based refinement}, \hypertarget{FR7}{(FR7) support export and import of results for replication, reuse, and sharing}, \hypertarget{FR8}{(FR8) persist user interactions and history across sessions}, and \hypertarget{FR9}{(FR9) provide documentation and developer support}. Non‑functionally, the dashboard emphasizes \hypertarget{NFR1}{(NFR1) usability through a responsive interface}, \hypertarget{NFR2}{(NFR2) transparency and reproducibility of queries and interactions}, \hypertarget{NFR3}{(NFR3) extensibility for new LR datasets}, competency questions, and LLMs, \hypertarget{NFR4}{(NFR4) interoperability with external services}, \hypertarget{NFR5}{(NFR5) reliability in retrieving structured LR data}, and \hypertarget{NFR6}{(NFR6) maintainability through modular component libraries and comprehensive API specifications}. Below, we describe how the frontend and backend components implement the requirements in their architecture.~\vspace{0.1cm}

\textbf{Frontend.} Through a responsive web interface, researchers can explore, synthesize, and reuse LR data by selecting curated competency questions (\hyperlink{FR1}{\textcolor{link}{FR1}}), posing custom ones (\hyperlink{FR2}{\textcolor{link}{FR2}}), interacting with the LLM and manually refining the dynamically generated visualizations, interpretations, and explanations (\hyperlink{FR5}{\textcolor{link}{FR5}}, \hyperlink{FR6}{\textcolor{link}{FR6}}) (cf. Section~\ref{sec:workflows}). The frontend integrates a \textit{workflow engine} that orchestrates the competency question-based pipeline in five steps: (1)  Selecting curated competency questions and their corresponding SPARQL queries (\hyperlink{FR1}{\textcolor{link}{FR1}}), or analyzing custom natural language competency questions and iteratively generating SPARQL queries (\hyperlink{FR2}{\textcolor{link}{FR2}}); (2) Executing SPARQL queries against the ORKG (\hyperlink{FR3}{\textcolor{link}{FR3}}); (3) Processing the retrieved data (\hyperlink{FR4}{\textcolor{link}{FR4}}); (4) Generating interactive visualizations enriched with interpretations and explanations (\hyperlink{FR4}{\textcolor{link}{FR4}}, \hyperlink{FR5}{\textcolor{link}{FR5}}); and (5) Supporting manual and prompt-based refinement of the results (\hyperlink{FR6}{\textcolor{link}{FR6}}). In addition, the frontend provides a \textit{history manager} for comprehensive control over LLM memory and context, offering detailed history for selective restoration, contextual memory management to retain or discard information, and cross‑session persistence to maintain progress across browser sessions (\hyperlink{FR8}{\textcolor{link}{FR8}}). The system also supports exporting the underlying data retrieved for curated competency questions (\hyperlink{FR7}{\textcolor{link}{FR7}}), as well as exporting and importing the complete content of custom competency questions (\hyperlink{FR7}{\textcolor{link}{FR7}}), facilitating replication, reuse, and sharing. Together, these components ensure that users can iteratively refine queries, visualizations, interpretations, and explanations in a transparent and reproducible manner (\hyperlink{NFR2}{\textcolor{link}{NFR2}}). Built with modern web technologies, \href{https://react.dev/}{\textit{\textcolor{link}{React}}}, \href{https://vite.dev/}{\textit{\textcolor{link}{Vite}}}, \href{https://www.typescriptlang.org/}{\textit{\textcolor{link}{TypeScript}}}, and \href{https://mui.com/material-ui/}{\textit{\textcolor{link}{Material‑UI}}}, and a \href{https://storybook.js.org/}{\textit{\textcolor{link}{Storybook}}}-based \href{https://empire-compass-storybook.tib.eu/?path=/docs/layout-dashboard--docs}{\textit{\textcolor{link}{component library}}} for consistent user interface and user experience design (\hyperlink{FR9}{\textcolor{link}{FR9}}), the frontend prioritizes usability (\hyperlink{NFR1}{\textcolor{link}{NFR1}}) and maintainability (\hyperlink{NFR6}{\textcolor{link}{NFR6}}). These qualities ensure effective interaction, reliable presentation of LR data, and adaptability for future developments.~\vspace{0.1cm}

\textbf{Backend.} The backend consists of the symbolic layer and the neural layer. The symbolic layer connects the dashboard to the ORKG via SPARQL queries (\hyperlink{FR3}{\textcolor{link}{FR3}}), ensuring transparency, reproducibility, and reliability in retrieving structured LR data and contextual information (\hyperlink{NFR2}{\textcolor{link}{NFR2}}, \hyperlink{NFR5}{\textcolor{link}{NFR5}}). In particular, it implements the \textit{competency question framework}, which comprises pairs of curated natural language competency questions investigated by an LR and the corresponding SPARQL queries to retrieve the required data (\hyperlink{FR1}{\textcolor{link}{FR1}}). For the KG-EmpiRE use case, the framework contains 16 pairs\footnote{We systematically derived and validated these 16 competency questions from literature in our previous work~\cite{Karras.2023}.}, while the NLP4RE ID Card use case provides 10 pairs\footnote{We collected these 10 competency questions directly from domain experts, namely the authors of the NLP4RE ID Card~\cite{Abualhaija.2024}.}. Each use case relies on its own graph schema\footnote{Graph schema of the \href{https://empire-compass.tib.eu/R186491/schema}{\textit{\textcolor{link}{KG-EmpiRE}}} use case and the \href{https://empire-compass.tib.eu/R1544125/schema}{\textit{\textcolor{link}{NLP4RE ID Card}}} use case.} to ensure semantic consistency (NFR2). The neural layer integrates LLMs from leading providers (\href{https://openai.com/}{\textit{\textcolor{link}{OpenAI}}} for ChatGPT models, \href{https://groq.com/}{\textit{\textcolor{link}{Groq}}} for LLaMA and DeepSeek models, \href{https://mistral.ai/}{\textit{\textcolor{link}{Mistral AI}}} for Mistral models, and \href{https://gemini.google/us/about/?hl=en}{\textit{\textcolor{link}{Google Generative AI}}} for Gemini and Gemma models). These LLMs generate SPARQL queries from custom natural language competency questions based on the respective graph schema (\hyperlink{FR2}{\textcolor{link}{FR2}}), interpret retrieved data (\hyperlink{FR3}{\textcolor{link}{FR3}}, \hyperlink{FR4}{\textcolor{link}{FR4}}), and provide visualizations enriched with interpretations and explanations tailored to the users' needs (\hyperlink{FR4}{\textcolor{link}{FR4}}, \hyperlink{FR5}{\textcolor{link}{FR5}}, \hyperlink{FR6}{\textcolor{link}{FR6}}). After login, users can use the system’s default LLM (GPT-4o mini) or select their preferred LLM (\hyperlink{NFR3}{\textcolor{link}{NFR3}}).
Together, both layers balance semantic accuracy with adaptive flexibility (\hyperlink{NFR1}{\textcolor{link}{NFR1}}, \hyperlink{NFR2}{\textcolor{link}{NFR2}}). In addition, the backend also manages the real‑time ORKG access (\hyperlink{FR3}{\textcolor{link}{FR3}}), the persistence of user interactions in a \href{https://firebase.google.com/products/firestore?hl=en}{\textit{\textcolor{link}{Firebase Firestore}}} (\hyperlink{FR8}{\textcolor{link}{FR8}}), automated \href{https://empire-compass.tib.eu/R186491/statistics}{\textit{\textcolor{link}{statistics}}} via Python scripts (\hyperlink{FR9}{\textcolor{link}{FR9}}), and a comprehensive \href{https://swagger.io/}{\textit{\textcolor{link}{Swagger}}}-based \href{https://empire-compass-backend.tib.eu/api-docs/}{\textit{\textcolor{link}{API documentation}}} for an interactive, machine-readable specification of all backend endpoints (\hyperlink{FR9}{\textcolor{link}{FR9}}, \hyperlink{NFR6}{\textcolor{link}{NFR6}}). The Swagger interface enables developers to explore request/response structures, test queries directly, and seamlessly connect to external services (\hyperlink{NFR4}{\textcolor{link}{NFR4}}). Overall, the backend prioritizes extensibility (\hyperlink{NFR3}{\textcolor{link}{NFR3}}), interoperability (\hyperlink{NFR4}{\textcolor{link}{NFR4}}), and maintainability (\hyperlink{NFR6}{\textcolor{link}{NFR6}}), enabling integration of new LR datasets, competency questions, LLMs, and external services.

\subsection{Workflows and User Interaction}\label{sec:workflows}
As shown in \figurename{~\ref{fig:architecture_workflows}, the dashboard supports two main workflows: Answering \textit{curated competency questions} (\href{https://empire-compass.tib.eu/R186491/questions/1}{\textit{\textcolor{link}{Workflow 1}}}) and answering \textit{custom competency questions} (\href{https://empire-compass.tib.eu/R186491/dynamic-question}{\textit{\textcolor{link}{Workflow 2}}}).\vspace{0.1cm}

\textbf{Workflow 1: Curated Competency Question.} Users select a curated competency question, which executes the corresponding SPARQL query against the ORKG. The retrieved data is processed and an interactive visualization is rendered. The visualization is enriched with pre-formulated interpretations and explanations\footnote{For curated questions, interpretations and explanations are currently written manually to guarantee correctness. In the long term, LLM‑generated ones will be adopted once their accuracy and trustworthiness have been validated (cf. Section~\ref{sec:evaluation_plan}).}. Users can explore results further and prompt the LLM in the context of the selected competency question, i.e., to obtain additional interpretations, explanations, and visualizations. The underlying data can be exported for replication and reuse. 

\textbf{Workflow 2: Custom Competency Question.} Users ask a natural language competency question tailored to their analysis needs. The selected LLM generates a SPARQL query consistent with the corresponding graph schema, which can be iteratively refined manually and prompt-based. The query is executed against the ORKG, and the retrieved data is processed into an interactive visualization. Interpretations and explanations are generated by the LLM to enrich the visualization\footnote{For custom questions, interpretations and explanations are LLM‑generated and remain subject to validation and refinement.}, while users can further edit outputs manually or by prompting the LLM. The complete content of the custom competency questions, including prompts, queries, visualizations, interpretations, explanations, and the underlying data, can be exported for replication, reuse, and sharing, and imported to resume analyses.~\vspace{0.1cm}

Both workflows follow a common pipeline of querying the ORKG, processing retrieved data, rendering interactive visualizations, providing interpretations and explanations, and enabling manual and prompt-based iterative refinement. Export and import functionality ensures replication, reuse, sharing, and continuity across sessions, with detailed history tracking for transparency and reproducibility.

\section{Evaluation Plan}\label{sec:evaluation_plan}
We organized the evaluation plan for \href{https://empire-compass.tib.eu/R186491/}{\textit{\textcolor{link}{EmpiRE-Compass}}} into short‑, mid‑, and long‑term actions to cover all aspects from immediate validation to large-scale evaluation. These actions are stepwise aligned with the tool’s central objectives of knowledge \textit{exploration}, \textit{synthesis}, and \textit{reuse}, clarifying which activities can be conducted directly and which require further development and community involvement.\vspace{0.1cm}

The \textbf{short‑term actions} focus on content‑wise \textit{exploration} of the two use cases (KG‑EmpiRE and NLP4RE ID Card). We will examine whether the curated competency questions, visualizations, interpretations, and explanations provided by the dashboard are relevant, correct, and useful for their respective fields. These evaluations will involve expert reviews with researchers in empirical RE and NLP4RE, using surveys similar to our prior work in engineering sciences~\cite{Karras.2024e}. For these actions, we plan to benefit from the interactive setting of the posters and tools session at REFSQ’26 and the co-located NLP4RE workshop to attract domain experts as participants.

The \textbf{mid‑term actions} will conduct broader pilot studies for both use cases. Participants will be asked to perform \textit{synthesis} tasks by exploring LR data and combining evidence for their custom competency questions. We will measure task completion time, coverage of relevant information, and user experience. For these actions, we plan to again benefit from the interactive setting at REFSQ’26 to attract participants, gather custom competency questions, and collect initial feedback from the conference attendees. These inputs will guide the design of the long‑term actions and help identify participants for follow-up studies.

The \textbf{long‑term actions} will address the large-scale evaluation of knowledge \textit{reuse} for both use cases. We will systematically examine the performance of LLMs generating visualizations, interpretations, and explanations by reusing the knowledge to answer the collected custom competency questions. This evaluation requires numerous custom competency questions and answers, and thus significant community support to ensure that the assessment reflects genuine research interests rather than author‑defined perspectives. We will combine quantitative metrics, such as completeness, accuracy, precision, and recall collected with automated approaches, with qualitative insights from domain experts gathered through focus groups and surveys.\vspace{0.1cm}

This staged evaluation plan enables a systematic assessment of \href{https://empire-compass.tib.eu/R186491/}{\textit{\textcolor{link}{EmpiRE-Compass}}}, by addressing immediate content exploration, intermediate piloting of the synthesis capabilities, and large-scale evaluation of knowledge reuse. By leveraging diverse studies and community involvement, we aim to demonstrate the technical robustness, user-friendliness, and scientific value of the \href{https://empire-compass.tib.eu/R186491/}{\textit{\textcolor{link}{EmpiRE-Compass}}}.

\section{Conclusion and Future Work}
\href{https://empire-compass.tib.eu/R186491/}{\textit{\textcolor{link}{EmpiRE-Compass}}} demonstrates how the integration of a neuro-symbolic approach into a dashboard can make LR data more transparent, reusable, and sustainable. By unifying the symbolic representation of semantically structured data in RKGs with the neural, generative capabilities of LLMs, researchers can explore, synthesize, and reuse knowledge without requiring advanced technical skills. The dashboard provides a modular system design with clearly defined workflows for curated and custom competency questions, ensuring that users can interact with LR data easily and dynamically while maintaining transparency and reproducibility. In addition, releasing \href{https://empire-compass.tib.eu/R186491/}{\textit{\textcolor{link}{EmpiRE-Compass}}} as an \href{https://github.com/okarras/EmpiRE-Compass}{\textit{\textcolor{link}{open‑source project}}} fosters reuse, adoption, and extension by the broader community. This openness ensures that the dashboard can evolve beyond the presented use cases and be adapted to diverse research domains.

Beyond the evaluation plan, we plan two main actions for future work. First, we aim to generalize the dashboard into a technical framework that can be applied to arbitrary RKGs, enabling other communities to set up their own instances for knowledge exploration, synthesis, and reuse. Second, we will extend the synergy between LLMs and RKGs by incorporating a complementary perspective in our upcoming research project \href{https://www.oliver-karras.de/portfolio/scid-quest/}{\textit{\textcolor{link}{SciD‑QuESt}}}. So far, \href{https://empire-compass.tib.eu/R186491/}{\textit{\textcolor{link}{EmpiRE-Compass}}} focuses on the use of LLMs to improve access to and reuse of RKG data. In \href{https://www.oliver-karras.de/portfolio/scid-quest/}{\textit{\textcolor{link}{SciD‑QuESt}}}, we will extend the dashboard so that users can contribute new, semantically structured data to the ORKG. Based on our two use cases, we will develop a questionnaire-based approach that empowers researchers to extract scientific knowledge from publications, with LLMs suggesting extraction candidates and humans validating them before import into the ORKG. This human‑in‑the‑loop workflow lowers the barrier to contributing FAIR (meta-)data while ensuring quality and trustworthiness. Together, these actions will allow \href{https://empire-compass.tib.eu/R186491/}{\textit{\textcolor{link}{EmpiRE-Compass}}} to fully exploit the synergistic potential of LLMs and RKGs, laying the foundation for sustainable LRs that can be collaborative, continuously updated, and reused, thereby ensuring the quality, reliability, and timeliness of research results from LRs in RE, SE, and beyond.

%\newpage 
% \begin{acknowledgments}
%   The work was funded by the German Research Foundation (DFG) 501865131 (NFDI4Energy) and 442146713 (NFDI4Ing) within the German National Research Data Infrastructure (NFDI).
% \end{acknowledgments}

%% The declaration on generative AI comes in effect
%% https://ceur-ws.org/GenAI/Policy.html
%% Taxonomy: https://ceur-ws.org/GenAI/Taxonomy.html
\section*{Declaration on Generative AI}
During the preparation of this work, the author(s) used Microsoft Copilot and Grammarly in order to: Grammar and spelling check. After using these tools/services, the authors reviewed and edited the content as needed and take full responsibility for the publication’s content. 

%%
%% Define the bibliography file to be used
\bibliography{sample-ceur-no-dois}

\end{document}